\documentclass[11pt]{article}
\usepackage{moriond,epsfig}

\def\Journal#1#2#3#4{{#1} {\bf #2}, #3 (#4)}

\def\be{\begin{equation}}
\def\ee{\end{equation}}
\def\bea{\begin{eqnarray}}
\def\eea{\end{eqnarray}}

\newcommand{\etal}{{\it et al}}
\newcommand{\bb}{{\bf b}}
\newcommand{\C}{{\bf C}}
\begin{document}
\vspace*{4cm}
\title{REDSHIFT DISTORTIONS AND CLUSTERING IN THE PSCZ SURVEY}

\author{ W.E. BALLINGER}

\secondaddress{Department of Physics, Astrophysics, Keble Road,\\
Oxford OX1 3RH, England}

\author{ A.N. TAYLOR, A.F. HEAVENS}

\secondaddress{Institute for Astronomy, University of Edinburgh,
Blackford Hill, Edinburgh EH9 3HJ, Scotland}

\author{ H. TADROS}

\secondaddress{Department of Physics, Astrophysics, Keble Road,\\
Oxford OX1 3RH, England}

\maketitle\abstracts{
We have constrained the redshift-distortion parameter $\beta \equiv \Omega^{0.6}/b$ and the real-space
power spectrum of the IRAS PSCz survey using a spherical-harmonic
redshift-distortion analysis combined with a data compression method which
is designed to deal with correlated parameters. Our latest result,
$\beta=0.4 \pm 0.1$, 
strongly rules out $\beta=1$.}

\section{Introduction}

It has been understood for a long time that redshift surveys are
systematically distorted by peculiar velocities~\cite{ja} --
so called {\em redshift distortions}, which can cause the observed
distribution of galaxies to become anisotropic. On large scales where linear
theory applies, redshift distortions are characterised by the parameter
\footnote{$\Omega_{\rm m}$ is the
contribution to the density parameter from matter. The bias parameter $b$ depends
on galaxy type, in this paper it refers to IRAS galaxies.} 
$\beta \equiv \Omega_{\rm m}^{0.6}/b$  -- see
Hamilton~\cite{ha98} for a detailed review.

In a classic paper, Kaiser~\cite{ka} derived a simple formula for the
effect of linear redshift distortions for a volume limited survey that
subtends a small opening angle at the observer (the ``distant
observer'' approximation -- all lines of sight are treated as
parallel). He showed that power is boosted 
by a factor $(1+\beta \mu^2_k)^2$, where $\mu^{ }_{k}$ is the cosine
of the angle between wavevector and line of
sight. Hamilton~\cite{ha92}~\cite{ha93} and Cole, Fisher \& Weinberg~\cite{cfw94}~\cite{cfw95} analysed all-sky IRAS surveys using a
method based on the Kaiser formalism, but were forced to break the
survey up into sections because of the constraints of the distant
observer approximation, losing information about the largest (and most
reliably linear) scales.

Rather than fit a square peg into a round hole, Fisher
\etal~\cite{fsl} and Heavens \& Taylor~\cite{ht} (HT) dropped the plane parallel approximation and used a spherical
harmonic decomposition to match the spherical nature of the IRAS
redshift surveys. HT used spherical Bessel functions to
decompose the density field radially; using eigenfunctions of the
Laplacian retains all the advantages of Fourier analysis. HT analysed
the IRAS 1.2Jy survey, fitting
$\beta$ and the amplitude of the power spectrum; Ballinger, Heavens \&
Taylor~\cite{bht}(BHT) extended this analysis to fit the shape of the power
spectrum. Tadros \etal~\cite{t} (T99) applied these techniques to the IRAS PSCz
survey -- section \ref{previous} includes a review of those results.

\section{Spherical Harmonic Formalism}

We will briefly review the formalism for
 spherical harmonic analysis -- see HT and T99 for details. The density field of the galaxy distribution
 $\rho({\bf s})$
is expanded in terms of spherical harmonics, $Y_{\ell m}$, and
a discrete set of spherical Bessel functions, $j_{\ell}$,
\begin{equation}
	\hat{\rho}_{\ell mn} = c_{\ell n} \int \! d^3s \, 
	\rho({\bf s}) w(s)j_{\ell}\left(k_{\ell n}
	s\right)
	Y^{*}_{\ell m}\left(\theta,\phi\right),
\label{transequ}
\end{equation}
where the $c_{\ell n}$ are normalization constants and $k_{ln}$ are
discrete wavenumbers.

These observed coefficients can be related to those of the true underlying
density field ($\delta_{\ell mn}$) by:
\be
	\hat{\rho}_{\ell mn} = \left(\rho_{0}\right)_{\ell mn} +
	\sum_{\ell' \!m'\! n' } \sum_{n'' }
        S^{nn''}
	W_{\ell \ell'}^{mm'}\left(\Phi^{n'' n'}_{\ell \ell'} +
	\beta V^{n'' n'}_{\ell \ell'}\right)
	\delta_{\ell'\!m'\!n'}.
\ee
The transition matrices ${\bf W}$, ${\bf \Phi}$, ${\bf V}$ and ${\bf S}$ describe the effects 
of the sky mask, the radial selection function, the linear redshift 
space distortion and the small scale distortion correction
respectively; $\left(\rho_{0}\right)_{\ell mn}$ is a mean term,
non-zero because of partial sky coverage.  The transition matrices are  derived and 
defined in HT and T99.

A likelihood approach is used to constrain $\beta$ and the real-space
power spectrum:
\be
-2 \ln {\cal L} [ {\bf D} | \beta , P(k)] = \ln ( \det {\bf C} ) +
{\bf D} {\bf C}^{-1} {\bf D}^{T},
\ee
where ${\bf C} = < {\bf D D}^{T} >$ and elements of the data vector
${\bf D}$ are given by $D_{\ell mn} \equiv [\hat\rho_{\ell mn}-(\rho_0)_{\ell mn}]/\bar\rho$ where
$\bar\rho$ is the mean number density -- Gaussian statistics are assumed. Two different parameterisations
of $P(k)$ are used: following HT a fixed shape is assumed and the
amplitude is fitted, following BHT a stepwise maximum likelihood
method is used, allowing the power spectrum to assume any shape
(within bin discreteness).

\section{Data -- The IRAS PSCz Survey}

The PSCz survey~\cite{sa} is a redshift catalogue complete down to 0.6Jy over $\sim 83
\% $ of the sky with a total of $\sim 15,000$ redshifts -- it is the largest all sky survey in
existence. As our method is
very precise,  we minimise systematic errors by using a more conservative flux cut (0.75Jy) and sky mask reducing the number of
redshifts by a factor of roughly two -- see T99.

\section{Previous Results}
\label{previous}

Both the fixed amplitude (HT) and stepwise $P(k)$ (BHT) methods were
applied to the PSCz by T99 for modes with $k \leq 0.13 h$ Mpc$^{-1}$. The first method produced $\beta= 0.58\pm 0.26$ and the amplitude
of the real space power measured at wavenumber $k=0.1h$ Mpc$^{-1}$
of $\Delta_{0.1}=0.42 \pm 0.03$ -- see Fig. 2 (dashed contours). Freeing the shape of the power
spectrum we find the consistent results $\beta=0.47\pm 0.16$ (conditional error), and
$\Delta_{0.1}=0.47 \pm 0.03$ -- Fig. 1. T99 also carried out
extensive tests on simulations, and the methods were found to be
reliable. In addition, we carried out a suite of tests for systematics effects 
        in the data and found the cut catalogue gave consistent results.

\begin{figure}
\begin{center}
\leavevmode
\epsfysize=2.3in
\epsfbox[25 180 465 510]{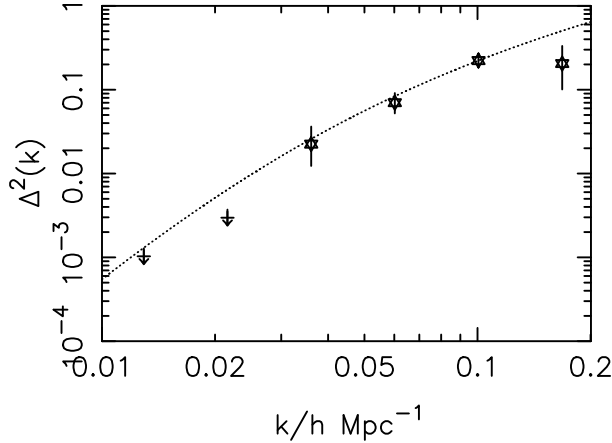}
\end{center}
\caption{The real-space power spectrum of the PSCz redshift survey in dimensionless units 
(from T99). 
The curve is a CDM model with $\Gamma = 0.2$.  The estimated 
value of $\beta$ is $0.47 \pm 0.16$ (conditional error).
\label{fig:P(k)}}
\end{figure}

The method was restricted to a small range of wavenumber because the
likelihood analysis involves the repeated inversion of a large $n
\times n$ matrix;
the time required for this process grows as $n^3$. More importantly,
the matrix rapidly became numerically unstable. The results above do
not strongly rule out either high ($\sim 1$) or low ($\sim0.5$) values
of $\beta$. It would be nice to overcome the matrix problem, extend the k-range and reduce the error bar.

\section{Data Compression}

It is possible to transform the (length $n$) data vector ${\bf D}$ to create a new,
smaller dataset, while retaining most of the information about the
parameters of interest~\cite{tth}. A new dataset is constructed which is a linear
combination of the original:
\be
{\bf D'} = {\bf B} {\bf D},
\ee
where ${\bf D'}$ is a new data vector of length $n' < n$, with a
corresponding transformation for the covariance matrix. 
TTH show that the optimal matrix ${\bf B}$ which minimises the
conditional error on a single parameter is made up from eigenvectors
of \footnote{This, like the Wiener filter, is
one of those marvellous data-handling methods which were invented by
astronomers then sent back in time so that signal processing
engineers could use them for the past fifty years.} 
\be
\label{eigen}
\C,_i\bb = \lambda \C\bb
\ee
where the comma refers to a derivative with respect to parameter
$i$. The new covariance matrix is smaller and also close to diagonal
-- hence much more stable. 

TTH suggested extending this to multi-parameter problems by
constructing a separate matrix ${\bf B}$ for each parameter and then
using singular value decomposition to combine the matrices
efficiently. However, Ballinger~\cite{b97} (see also Taylor
\etal~\cite{thbt}) tested this and found that error ellipses/ellipsoids
tended to grow along correlation axes -- only conditional errors are
constrained. Instead it was shown that equation (\ref{eigen}) could be
used to constrain the marginal errors by using one or more linear
combinations of the original parameters which lie directly along the
correlation axes.

Unfortunately solving equation (\ref{eigen}) involves manipulating
uncompressed $n \times n$ matrices. This usually isn't a timescale
problem -- it need only be done once -- but still suffers from
numerical difficulties. To avoid this problem, we split the data into
several sections and compressed them separately, while still retaining
full information about correlations between modes in different
sections. Strictly speaking this is slightly less optimal than
compressing the whole dataset in one go, but it should make a
negligible difference in practice and will not bias the result.

\section{New Results and conclusions}
Fig. 2 shows the old $k_{\rm max}=0.13h$ Mpc$^{-1}$ result together with the new
$k_{\rm max}=0.2h$ Mpc$^{-1}$ obtained using data compression. The original 4644
modes were compressed down to 2278. The new, lower,
value of $\beta = 0.4 \pm 0.1$ is consistent with the previous result but the error
ellipse is considerably smaller; the best-fit amplitude of the power spectrum
is essentially unchanged.  $\beta = 1$, corresponding to $\Omega_{\rm m}=1$, $b=1$, is
now strongly disfavoured. More details will be in Ballinger \etal\ and
Taylor \etal\ (in preparation).

The low value of $\beta$ is consistent with
currently popular cosmological models with a low value of $\Omega_{\rm
matter}$ if the IRAS bias parameter is close to unity. The value is
somewhat lower than that from other analyses of the PSCz survey~\cite{bran}~\cite{rr}~\cite{s1}~\cite{s2}, but
not inconsistent.
 It is
consistent with the recent {\em velocity-velocity} comparison results
from peculiar velocity catalogues~\cite{w97}, but somewhat lower
than the corresponding {\em density-density} value~\cite{sig}. See
Willick~\cite{w00} for a discussion.

\begin{figure}
\begin{center}
\leavevmode
\psfig{figure=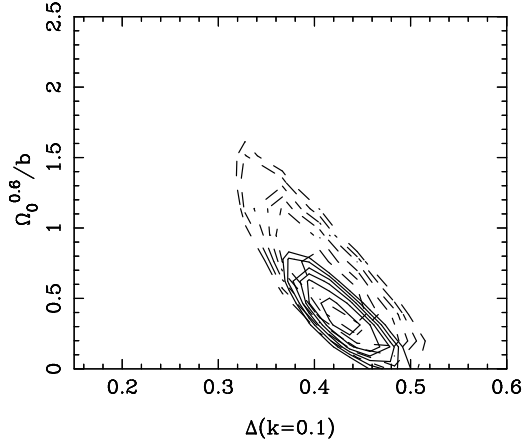,height=2.3in}
\caption{Likelihood contours for the two parameters
$\beta = \Omega_0^{0.6}/b$
and the amplitude of the power spectrum for
the PSCz survey. The new
data-compression $k_{\rm max}=0.2h$ Mpc$^{-1}$ results (solid contours) are plotted
along with the original $k_{\rm max}=0.13h$ Mpc$^{-1}$ results from T99 
(dashed contours). Contours
are plotted at intervals of $\delta \ln {\cal{L}} = 0.5.$
\label{fig:new2par}}
\end{center}
\end{figure}

\end{document}